\documentclass{procmult}
\usepackage{amsmath}
\usepackage{amsfonts}
\usepackage[dvips]{graphicx}

\begin{document}

\newcommand\be{\begin{equation}}
\newcommand\ee{\end{equation}}
\newcommand\bea{\begin{eqnarray}}
\newcommand\eea{\end{eqnarray}}
\newcommand\bseq{\begin{subequations}} 
\newcommand\eseq{\end{subequations}}
\newcommand\bcas{\begin{cases}}
\newcommand\ecas{\end{cases}}
\newcommand{\p}{\partial}
\newcommand{\f}{\frac}

\title{Bianchi IX in the GUP approach}
\author{Marco Valerio Battisti$^\dag$ \and Giovanni Montani$^\S$}
\institute{\small{
$^\dag$ ICRA and Centre de Physique Th\'eorique de Luminy, Universit\'e de la M\'editerran\'ee F-13288, Marseille EU, battisti@icra.it \\
$^\S$ ICRA, ICRANet, ENEA and Dipartimento di Fisica, Universit\`a di Roma ``Sapienza'' P.le A. Moro 5, 00185 Rome EU, montani@icra.it} 
}

\maketitle

\abstract{The Bianchi IX cosmological model (through Bianchi I and II) is analyzed in the framework of a generalized uncertainty principle. In particular, the anisotropies of the Universe are described by a deformed Heisenberg algebra. Three main results are in order. (i) The Universe can not isotropize because of the deformed Kasner dynamics. (ii) The triangular allowed domain is asymptotically stationary with respect to the particle (Universe) and its bounces against the walls are not interrupted by the deformed effects. (iii) No reflection law can be in obtained since the Bianchi II model is no longer analytically integrable.}

\bigskip

The existence of a fundamental scale, by which the continuum space-time picture that we have used from our experience at large scales probably breaks down, may be taken as a general feature of any quantum theory of gravity (for a review see \cite{gar}). This claim can be formalized modifying the canonical uncertainty principle as (we adopt units such that $\hbar=c=16\pi G=1$) 
\be\label{gup}
\Delta q \Delta p\geq \f 1 2\left(1+\beta (\Delta p)^2+\delta\right),
\ee  
where $\beta$ and $\delta$ are positive deformation parameters. This is the so-called generalized uncertainty principle (GUP) which appeared in string theory \cite{String}, considerations on the proprieties of black holes \cite{Mag} and de Sitter space \cite{Sny}. From the string theory point of view, the relation above is a consequence of the fact that strings can not probe distances below the string scale. The GUP (\ref{gup}) implies a finite minimal uncertainty in the position $\Delta q_0=\sqrt\beta$ and can be recovered by deforming the canonical commutation relations as $[q,p]=i(1+\beta p^2)$ as soon as $\delta=\beta \langle p\rangle^2$. Recently, the GUP framework has received notable interest and a wide work has been made on this field in a large variety of directions (see \cite{GUP} and references therein).

In this paper we describe the dynamics of the Bianchi cosmological models in the GUP framework reviewing the results of \cite{BM09}. In particular, we analyze the most general homogeneous model (Bianchi IX or Mixmaster) passing through the necessary steps of Bianchi I and II. The GUP approach has been previously implemented to the FRW model filled with a massless scalar field \cite{BM07a} as well as to the Taub Universe \cite{BM07b}. In the first case \cite{BM07a}, the big-bang singularity appears to be probabilistically removed but no evidences for a big-bounce, as predicted by the loop approach \cite{APS}, arise (a cosmological bounce \`a la loop quantum cosmology has been obtained from a deformed Heisenberg algebra in \cite{Bat}). The GUP Taub Universe \cite{BM07b} is also singularity-free and this feature is relevant since allows a phenomenological comparison with the polymer (loop) Taub Universe \cite{BLM08}. In fact, in the latter model, the cosmological singularity appears to be not removed. The analysis of the Bianchi models then improve such a research line since the two anisotropies of the Universe are now described by a deformed Heisenberg algebra.  

The Bianchi Universes are spatially homogeneous cosmological models (for reviews see \cite{rev}) and their dynamics is summarized in the scalar constraint which, in the Misner scheme, reads
\be\label{scacon} 
H=-p_\alpha^2+p_+^2+p_-^2+e^{4\alpha}V(\gamma_\pm)=0,
\ee
where the lapse function $N=N(t)$ has been fixed by the time gauge $\dot\alpha=1$ as $N=-e^{3\alpha}/2p_\alpha$. The variable $\alpha=\alpha(t)$ describes the isotropic expansion of the Universe while its shape changes (the anisotropies) are determinated via $\gamma_\pm=\gamma_\pm(t)$. Homogeneity reduces the phase space of general relativity to six dimensions. In the Hamiltonian framework the cosmological singularity appears for $\alpha\rightarrow-\infty$ and the differences between the Bianchi models are summarized in the potential term $V(\gamma_\pm)$ which is related to the three-dimensional scalar of curvature. As well-known, to describe the dynamical evolution of a system in general relativity a choice of time has to be performed. This can be basically accomplished in a relational way (with respect to an other field) or with respect to an internal time which is constructed from phase space variables. The ADM reduction of the dynamics relies on the idea to solve the scalar constraint with respect to a suitably chosen momentum. This way, an effective Hamiltonian which depends only on the physical degrees of freedom of the system is naturally recovered. Since the volume $\mathcal V$ of the Universe is $\mathcal V\propto e^{3\alpha}$, the variable $\alpha$ can be regarded as a good clock for the evolution and therefore the ADM picture arises as soon as the constraint (\ref{scacon}) is solved with respect to $p_\alpha$. Explicitly, we obtain
\be\label{admham}
-p_\alpha=\mathcal H=\left(p_+^2+p_-^2+e^{4\alpha}V(\gamma_\pm)\right)^{1/2},
\ee 
where $\mathcal H$ is a time-dependent Hamiltonian from which is possible to extract, for a given symplectic structure, all the dynamical informations about the homogeneous cosmological models. 

Let us now analyze the modifications induced on a $2n$-dimensional phase space by the GUP framework. Assuming the translation group as not deformed, i.e. $[p_i,p_j]=0$, and the existence of a new deformation parameter $\beta'>0$, the phase space algebra is the one in which the fundamental Poisson brackets are \cite{Ben} 
\bea\label{defal}
\{q_i,p_j\}&=&\delta_{ij}(1+\beta p^2)+\beta'p_ip_j,\\ \nonumber
\{p_i,p_j\}&=&0,\\ \nonumber
\{q_i,q_j\}&=&\f{(2\beta-\beta')+(2\beta+\beta')\beta p^2}{1+\beta p^2}(p_iq_j-p_jq_i).
\eea  
These relations are obtained assuming that $\beta$ and $\beta'$ are independent constants with respect to $\hbar$. From a string theory point of view, keeping the parameters $\beta$ and $\beta'$ fixed as $\hbar\rightarrow0$ corresponds to keeping the string momentum scale fixed while the string length scale shrinks to zero \cite{String}. The Poisson bracket for any phase space function can be straightforward obtained from (\ref{defal}) and reads
\be
\{F,G\}=\left(\f{\p F}{\p q_i}\f{\p G}{\p p_j}-\f{\p F}{\p p_i}\f{\p G}{\p q_j}\right)\{q_i,p_j\}+\f{\p F}{\p q_i}\f{\p G}{\p q_j}\{q_i,q_j\}.
\ee
It is worth noting that for $\beta'=2\beta$ the coordinates $q_i$ become commutative up to higher order corrections, i.e. $\{q_i,q_j\}=0+\mathcal O(\beta^2)$. This can be then considered a preferred choice of parameters and from now on we analyze this case. However, although we neglect terms like $\mathcal O(\beta^2)$, the case in which $\beta p^2\gg1$ is allowed since in such a framework no restrictions on the $p$-domain arise, i.e. $p\in\mathbb R$. The deformed classical dynamics of the Bianchi models can be therefore obtained from the symplectic algebra (\ref{defal}) for $\beta'=2\beta$. The time evolution of the anisotropies and momenta, with respect to the ADM Hamiltonian (\ref{admham}), is thus given by ($i,j=\pm$)
\bea\label{defeq}
\dot \gamma_i&=&\{\gamma_i,\mathcal H\}=\f1{\mathcal H}\left[(1+\beta p^2)\delta_{ij}+2\beta p_ip_j\right]p_j,\\ \nonumber
\dot p_i&=&\{p_i,\mathcal H\}=-\f{e^{4\alpha}}{2\mathcal H}\left[(1+\beta p^2)\delta_{ij}+2\beta p_ip_j\right]\f{\p V}{\p \gamma_j},
\eea 
where the dot denotes differentiation with respect to the time variable $\alpha$ and $p^2=p_+^2+p_-^2$. These are the deformed equations of motion for the homogeneous Universes and the ordinary ones are recovered in the $\beta=0$ case. 

The Bianchi I model is the most simpler homogeneous model and describes a Universe with flat space sections \cite{rev}. Its line element is invariant under the group of three-dimensional translations and the spatial Cauchy surfaces can be then identified with $\mathbb R^3$. This Universe contains as a special case the flat FRW model which is obtained as soon as the isotropy condition is taken into account. Bianchi I corresponds to the case $V(\gamma_\pm)=0$ and thus, from the Hamiltonian (\ref{admham}), it is described by a two-dimensional massless scalar relativistic particle. The velocity of the particle (Universe) is modified by the deformed symplectic geometry and, from the first equation of (\ref{defeq}), it reads
\be\label{anivel}
\dot\gamma^2=\dot\gamma_+^2+\dot\gamma_-^2=\f{p^2}{\mathcal H^2}\left(1+6\mu+9\mu^2\right)=1+6\mu+9\mu^2,
\ee 
where $\mu=\beta p^2$. For $\beta\rightarrow0$ ($\mu\ll1$), the standard Kasner velocity $\dot\gamma^2=1$ is recovered. Therefore, the effects of an anisotropies cut-off imply that the point-Universe moves faster than the ordinary case. In the deformed scheme the solution is still Kasner-like ($\dot\gamma_\pm=C_\pm(\beta),\dot p_\pm=0$), but this behavior is modified by the equation (\ref{anivel}). In particular, the second Kasner-relation between the Kasner indices $s_1, s_2, s_3$ appears to be deformed as 
\be\label{kasrel}
s_1^2+s_2^2+s_3^2=1+4\mu+6\mu^2,
\ee
while the first one $s_1+s_2+s_3=1$ remains unchanged. As usual, for $\beta=0$, the standard one is recovered. Two remarks are therefore in order. (i) The GUP acts in an opposite way with respect to a massless scalar field in the standard model. In that case the chaotic behavior of the Mixmaster Universe is tamed \cite{scafield}. On the other hand, in the GUP framework, all the terms on the right hand side of (\ref{kasrel}) are positive and it means that the Universe cannot isotropize, i.e. it can not reach the stage such that the Kasner indices are equal. (ii) For every non-zero $\mu$, two indices can be negative at the same time. Thus, as the volume of the Universe contracts toward the classical singularity, distances can shrink along one direction and grow along the other two. In the ordinary case the contraction is along two directions.

The natural bridge between Bianchi I and the Mixmaster Umiverse is represented by the Bianchi II model. Its dynamics is the one of a two-dimensional particle bouncing against a single wall and it corresponds to the Mixmaster dynamics when only one of the three equivalent potential walls is taken into account \cite{rev}. In the Hamiltonian framework, Bianchi II is described by the potential term $V(\gamma_\pm)=e^{-8\gamma_+}$ and this expression can be directly obtained from the one of Bianchi IX in an asymptotic region. The main features of Bianchi IX, as the BKL map, are obtained considering such a simplified model since it is, in the ordinary framework, an integrable system differently from Bianchi IX itself. The BKL map, which is as the basis of the analysis of the stochastic and chaotic proprieties of the Mixmaster Universe, appears to be the reflection law of the $\gamma$-particle (Universe) against the potential wall. A fundamental difference which arises in the deformed framework with respect to the ordinary one, is that $\mathcal H$ is no longer a constant of motion near the classical singularity. The wall-velocity $\dot\gamma_w$ is then modified as \cite{BM09}
\be\label{walvel}
\dot\gamma_w=\f1{36\mu}\left(-4+22^{1/3}g^{-1/3}+2^{2/3}g^{1/3}\right),
\ee
where $g=2+81\mu\dot\gamma^2+9\sqrt{\mu\dot\gamma^2(4+81\mu\dot\gamma^2)}$. From the two velocity equations (\ref{anivel}) and (\ref{walvel}), it is possible to discuss the details of the bounce. In the standard case the particle (Universe) moves twice as fast as the receding potential wall, independently of its momentum (namely its energy). In the deformed framework the particle-velocity, as well as the potential-velocity, depends on the anisotropy momentum and on the deformation parameter $\beta$. Also in this case the particle moves faster than the wall since the relation $\dot\gamma_w<\dot\gamma$ is always verified (see Fig. 1) and a bounce takes then place also in the deformed picture. Furthermore, in the asymptotic limit $\mu\gg1$ the maximum angle in order the bounce against the wall to occur is given by $|\theta_{\max}|=\pi/2$, differently from the ordinary case ($\dot\gamma_w/\dot\gamma=1/2$) where the maximum incidence angle is given by $|\theta_{\max}|=\pi/3$. 
\begin{figure}
\centering
\includegraphics[height=1.5in]{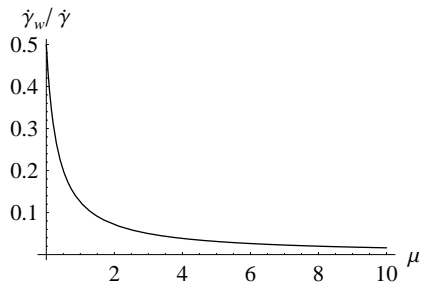}
\caption{The potential wall velocity $\dot\gamma_w$ with respect to the particle one $\dot\gamma$ in function of $\mu=\beta p^2$. In the $\mu\rightarrow0$ limit, the ordinary behavior $\dot\gamma_w/\dot\gamma=1/2$ is recovered.} 
\end{figure}
The $\gamma$-particle bounce against the wall is thus improved in the sense that no longer maximum limit angle appears. However, the main difference with respect to the ordinary picture is that the deformed Bianchi II model is not analytically integrable. No reflection map can be then in general inferred. In fact, it is no longer possible to find two constants of motion in the GUP picture. For more details see \cite{BM09}. 

On the basis of the previous analysis of the GUP Bianchi I and II models we know several features of the deformed Mixmaster Universe. We recall (see \cite{rev}) that the Bianchi IX geometry is invariant under the three-dimensional rotation group (this Universe is the generalization of the closed FRW model when the isotropy hypothesis is relaxed) and its potential term is given by $V(\gamma_\pm)=e^{4(\gamma_++\sqrt3\gamma_-)}+e^{4(\gamma_+-\sqrt3\gamma_-)}+e^{-8\gamma_+}$. The evolution of the Mixmaster Universe is that of a two-dimensional particle bouncing (the single bounce is described by the Bianchi II model) infinite times against three walls which rise steeply toward the singularity. Between two succeeding bounces the system is described by the Kasner evolution and the permutations of the expanding-contracting directions is given by the BKL map \cite{BKL} showing the chaotic behavior of such a dynamics \cite{chaos}. Two conclusions on the deformed Mixmaster Universe can be thus inferred \cite{BM09}. 
\begin{itemize}
	\item When the ultra-deformed regime is reached ($\mu\gg1$), i.e. when the $\gamma$-particle (Universe) has the momentum bigger than the cut-off one, the triangular closed domain appears to be stationary with respect to the particle itself. The bounces of the particle are then increased by the presence of deformation terms, i.e. by the non-zero minimal uncertainty in the anisotropies. 
	\item No BKL map (reflection law) can be in general obtained. It arises analyzing the single bounce against a given wall of the equilateral-triangular domain and the Bianchi II model is no longer an integrable system in the deformed picture. The chaotic behavior of the Bianchi IX model is then not tamed by GUP effects, i.e. the deformed Mixmaster Universe is still a chaotic system.
\end{itemize}  

As last point we stress the differences between our model and the loop Mixmaster dynamics \cite{Mixl}. In the loop Bianchi IX model the classical reflections of the $\gamma$-particle stop after a finite amount of time and the Mixmaster chaos is therefore suppressed. In this framework, although the analysis is performed through the ADM reduction of the dynamics as we did, all the three scale factors are quantized using the loop techniques. On the other hand, in our approach the time variable (related to the volume of the Universe) is treated in the standard way and only the two physical degrees of freedom of the Universe (the anisotropies) are considered as deformed.  

\bigskip

{\it Acknowledgments.} M. V. B. thanks ''Fondazione Angelo Della Riccia'' for financial support.

\end{document}